\documentclass[11pt]{article}

\usepackage[margin=1in]{geometry}
\usepackage{graphicx}
\usepackage{hyperref}
\usepackage{xcolor}
\usepackage{caption}
\usepackage{authblk}
\usepackage{microtype}
\usepackage{enumitem}
\usepackage{url}

\hypersetup{
  colorlinks=true,
  linkcolor=black,
  citecolor=black,
  urlcolor=blue!60!black,
  pdftitle={APTLAS: A Curated Literature Atlas for APT},
  pdfauthor={Bavley Guerguis and Nabil Bassim}
}

\title{\bfseries APTLAS:\\An Indexed APT Literature Repository}

\author[1,2]{Bavley Guerguis\thanks{Corresponding author: \texttt{guerguib@mcmaster.ca}}}
\author[1,2]{Nabil Bassim}
\affil[1]{\small Department of Materials Science and Engineering, McMaster University, Hamilton, ON L8S 4L7, Canada}
\affil[2]{\small Canadian Centre for Electron Microscopy, McMaster University, Hamilton, ON L8S 4M1, Canada}

\date{}
\begin{document}
\maketitle

\begin{abstract}
\noindent
Atom probe tomography (APT) literature is broad, rapidly growing, and dispersed across a wide range of journals, which can make it difficult to identify prior work on a given material system, instrument, or analytical approach. Conventional search engines (e.g., Google Scholar) excel at general retrieval but do not preserve the domain-specific metadata that often determines the relevance of an APT publication (e.g., analysis mode, laser wavelength, or instrument configuration). Herein, APTLAS is presented, which is an indexed repository of published APT literature. At present, the database contains $\sim$2,300 records, each accompanied by  metadata extracted from the source publication. The accompanying web tool, available at \url{https://aptlas.bavleyguerguis.com/}, allows users to browse and filter by material system, instrument, application, publication type, or keyword search. 
\end{abstract}

\section{Introduction}

Although the technique was historically confined to metallurgy, the commercialization of laser-pulsed instruments has facilitated the rapid expansion of atom probe tomography (APT) over the past two decades \cite{gault2021atom}. As a consequence, the corresponding literature has grown rapidly and is distributed across general-interest journals (e.g., \textit{Nature}, \textit{Science}), microscopy and characterization journals (e.g., \textit{Ultramicroscopy}, \textit{Microscopy and Microanalysis}), and a wide range of materials-specific journals (e.g., \textit{Acta Materialia}, \textit{Applied Physics Letters}, \textit{Journal of Nuclear Materials}).

There is a practical impediment for one attempting to identify prior work based on a specific combination of variables (e.g., laser-pulsed APT of boron-doped silicon on a LEAP 5000 XS). Conventional search engines, while powerful, are not designed to filter on domain-specific metadata, such as the instrument model or the analysis mode. This challenge is not unique to APT, but it is particularly pronounced given the high dimensionality of the relevant parameter space and the strong dependence of the measurement on the experimental conditions employed (i.e., compositional biases). Herein, we describe the construction and deployment of a curated index of APT literature. We outline the literature acquisition process, the categorization schema, the large language model (LLM)-based metadata extraction, and the web-based tool to navigate the database. 

\section{Methodology}

\begin{figure}[t]
  \centering
  \includegraphics[width=0.95\linewidth]{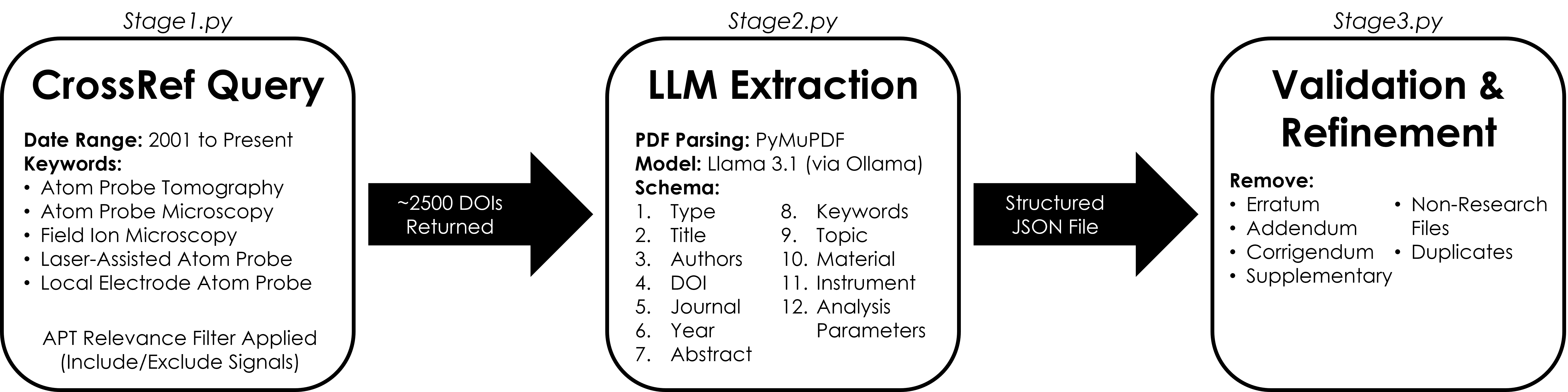}
  \caption{Overview of the APTLAS construction pipeline.}
  \label{fig:pipeline}
\end{figure}

The construction of APTLAS proceeded in three stages: (1) literature acquisition, (2) metadata extraction, and (3) validation/refinement. Each is briefly described below, with a schematic of the overall pipeline shown in Fig.~\ref{fig:pipeline}.

\subsection{Literature Acquisition}

First, a list of candidate publications was assembled by querying the CrossRef API with the keywords \textit{Atom Probe Tomography}, \textit{Atom Probe Microscopy}, \textit{Field Ion Microscopy}, \textit{Laser-Assisted Atom Probe}, and \textit{Local Electrode Atom Probe} (\texttt{Stage1.py}). The query was restricted to publications from 2001 onwards, though select foundational publications predating this range are also included. 

\subsection{Categorization Schema and LLM-Based Metadata Extraction}

A core design decision was the choice of categories under which each publication would be indexed. The intent was to balance granularity against tractability.

Categories are organized hierarchically and comprise of: (1) \textit{Publication type} (i.e., Journal Article, Letter, Review, Conference Proceeding, Thesis); (2) \textit{Topic} (i.e., Instrumentation and Technique Development, Specimen Preparation and Experimental Methodology, Data Reconstruction and Physical Modeling, Data Analysis and Quantification Methods, Computational and Machine-Learning Approaches, Correlative and Multimodal Characterization, and Materials Science and Application Studies);  (3) \textit{Material System}  (i.e., Metals and Alloys, Semiconductors, Ceramics and Oxides, Functional and Energy Materials, Thin Films and Nanostructures, Nuclear and Irradiated Materials, Geological and Extraterrestrial Materials, and Biomaterials and Organic Material); (4) \textit{Instrument}  (i.e., name, configuration, and laser wavelength, when applicable); and (5)  \textit{Analysis Conditions} (i.e., analysis mode, base temperature, pulse rate, detection rate, and the pulse fraction or laser pulse energy).

Each publication was passed to a local Llama 3.1 model served via Ollama with a structured-output prompt (\texttt{Stage2.py}) and instructed to return a JSON object conforming to the APTLAS schema, with each field populated either from the text or with an explicit \texttt{null} where the information was not reported.

\subsection{Validation and Assembly}

Extracted records were subsequently validated and refined (\texttt{Stage3.py}). Duplicates were identified by content hashing and DOI matching, and were removed. Non-research documents were also excluded, including errata, addenda, corrigenda, and supplementary data files. The remaining records were assembled into a single JSON file (\texttt{data.json}). Where information was unavailable or ambiguous, the corresponding field is set to \texttt{null} in the database.

\section{APTLAS}

\begin{figure}[t]
  \centering
  \includegraphics[width=0.95\linewidth]{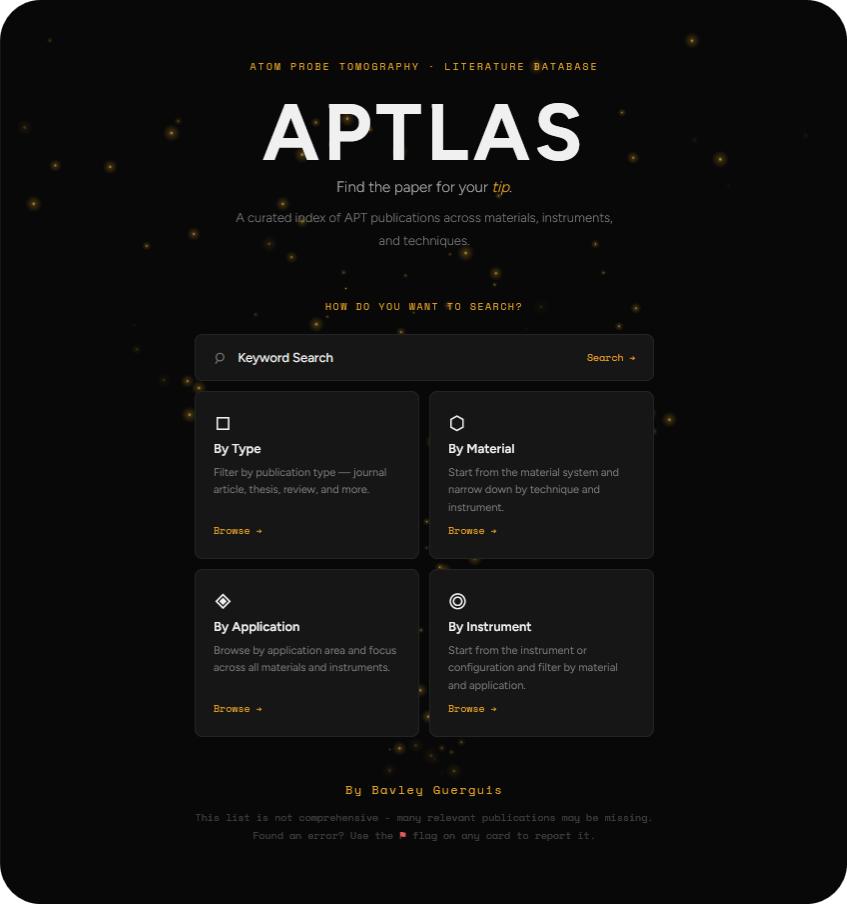}
  \caption{Landing page of the APTLAS web tool.}
  \label{fig:landing}
\end{figure}

The APTLAS tool is a single-page web application implemented in HTML. The tool is hosted at \url{https://aptlas.bavleyguerguis.com/}, and the source files can be downloaded and run locally by placing \texttt{index.html} and \texttt{data.json} in the same directory and opening the former in any browser. The landing page (Fig.~\ref{fig:landing}) presents five entry points: four browsing modes (\textit{By Type}, \textit{By Material}, \textit{By Application}, and \textit{By Instrument}) and keyword.

\begin{figure}[t]
  \centering
  \includegraphics[width=0.95\linewidth]{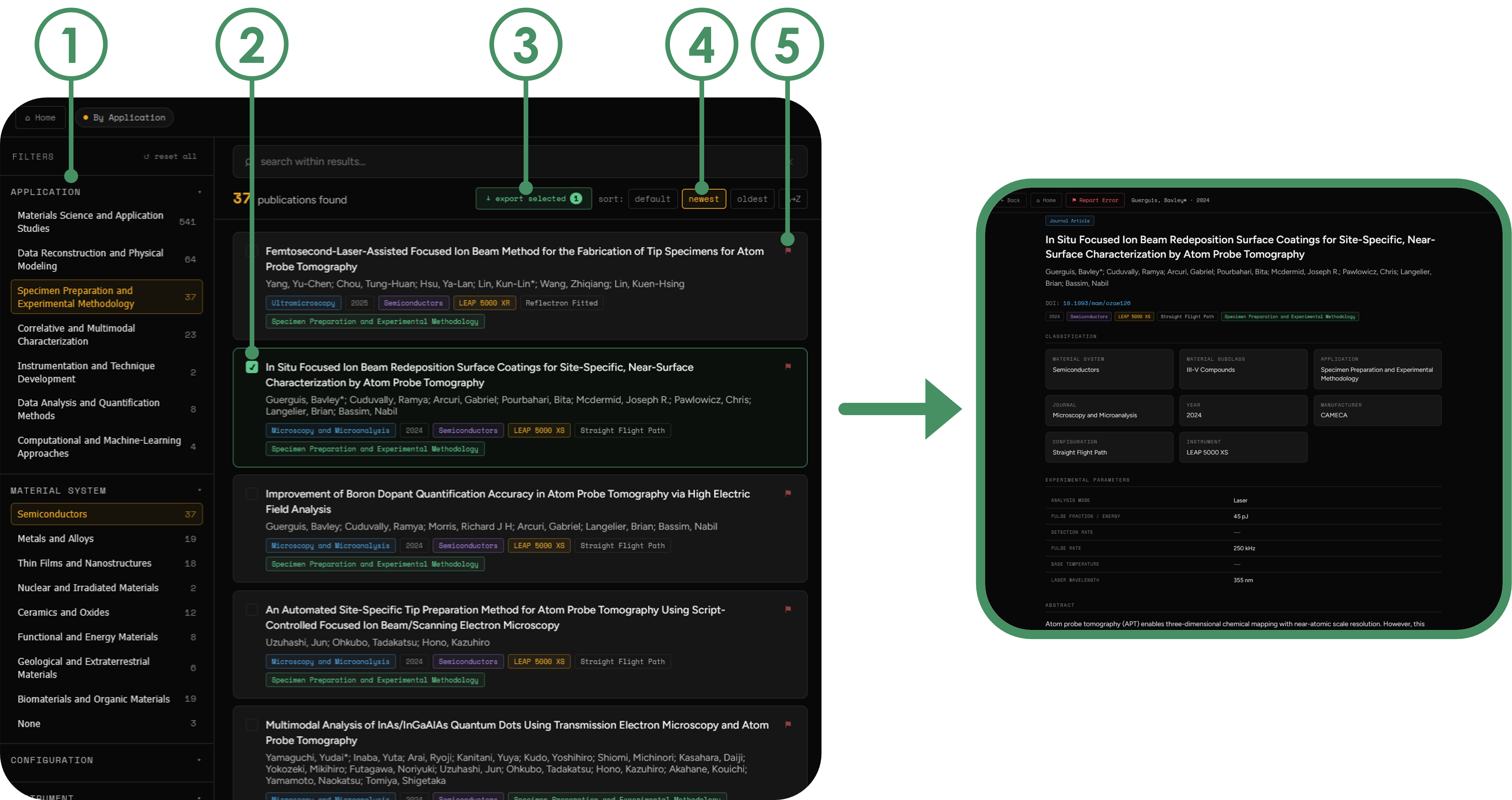}
  \caption{Filtered results view and the per-publication detail view.}
  \label{fig:detail}
\end{figure}

An example of a filtered results view is shown in Fig.~\ref{fig:detail}, with the key interactive elements labeled. The filter panel (1) allows the result set to be constrained by any combination of secondary filters (e.g., material system, instrument, configuration, topic, journal, and year range). Individual records can be selected using the per-card checkbox (2). Selected records can be exported as a plain-text citation list using the export button (3) (it is important to note that the export only contains simplified citation metadata). Results can also be ordered using the sort control (4), which supports sorting by relevance, year, or alphabetical order. Lastly, the report button (5) allows users to flag records that contain suspected extraction errors or incorrect metadata for review. 

Clicking any card opens a detailed view listing all extracted metadata for that publication, including the publicly available abstract and a hyperlink to the DOI. Where a field was not reported in the source publication, it is omitted.

\section{Limitations and Future Work}

Several limitations of the current implementation should be acknowledged. First, the LLM-based extraction, while fairly reliable, is not error-free (particularly the experimental parameters, where identififcation and correction of errors is an on-going effort). Users are encouraged to verify metadata against the original publication when accuracy is critical. Second, the schema, while extensive, is necessarily a simplification of a complex literature. Some publications do not fit cleanly into a single material system or topic, and the categorical assignment in such cases may be subjective (or erroneous).

Future work will focus on (i) periodic re-querying of CrossRef to incorporate newly published work, (ii) refinement of the extraction and the use of more capable models as they become available, and (iii) the assessment of more granular extraction targets, such as the specific dopant species investigated.

\section{Conclusion}

APTLAS is a curated index of the published APT literature, accompanied by a web tool for browsing and filtering. The tool is intended to reduce the friction associated with identifying prior work in a niche but rapidly growing technique.

\section*{Code and Data Availability}

The APTLAS web tool is available at \url{https://aptlas.bavleyguerguis.com/}. The two files comprising the tool --- \texttt{index.html} (the interface) and \texttt{data.json} (the database) --- can be downloaded from the website and run locally in any browser. The extraction pipeline scripts and the source files are also openly available via the project repository (please contact the corresponding author for the current link).

\section*{Acknowledgements}

 The authors are grateful to the many members of the APT community whose published work forms the basis of this resource.

\bibliographystyle{unsrt}

\end{document}